%% file: main.tex
\documentclass[sigconf]{acmart}

\usepackage{soul, color}
\usepackage{mdframed}
\usepackage{tikz}
\usepackage{subcaption}
\usetikzlibrary{calc,positioning,shapes.geometric,shapes.symbols,shapes.misc}
\usepackage{bm}

\AtBeginDocument{%
  \providecommand\BibTeX{{%
    \normalfont B\kern-0.5em{\scshape i\kern-0.25em b}\kern-0.8em\TeX}}}

\setcopyright{acmcopyright}
\copyrightyear{2023}
\acmYear{2023}
\acmDOI{XXXXXXX.XXXXXXX}

\acmConference[CONSEQUENCES '24]{Causality, Counterfactuals \& Sequential Decision-Making}{Oct. 14th-18th 2024}{Bari, Italy}

\acmBooktitle{CONSEQUENCES '24: Causality, Counterfactuals \& Sequential Decision-Making, Oct. 14th-18th 2024, Bari, Italy} 

\begin{document}

\title[The Importance of Causality in Decision Making]{The Importance of Causality in Decision Making: A Perspective on Recommender Systems}

\titlenote{Based on a full paper published at ACM Transaction on Recommender Systems \cite{cavenaghi2024towards}.}

\author{Emanuele Cavenaghi}
\email{ecavenaghi@unibz.it}
\orcid{0000-0002-0235-0421}
\affiliation{%
    \institution{Free University of Bozen-Bolzano}
    \streetaddress{Piazza Università, 1}
    \city{Bolzano}
    \country{Italy}
    \postcode{39100}
}
\author{Alessio Zanga}
\email{alessio.zanga@unimib.it}
\orcid{0000-0003-4423-2121}
\affiliation{%
    \institution{University of Milano-Bicocca, Milano, Italy}
    \streetaddress{Piazza dell'Ateneo Nuovo, 1}
    \city{}
    \country{}
    \postcode{20126}
}
\affiliation{%
    \institution{F. Hoffmann - La Roche Ltd, Basel, Switzerland}
    \streetaddress{Grenzacherstrasse 124}
    \city{}
    \country{}
    \postcode{4070}
}
\author{Fabio Stella}
\email{fabio.stella@unimib.it}
\orcid{0000-0002-1394-0507}
\affiliation{%
    \institution{University of Milano-Bicocca}
    \streetaddress{Piazza dell'Ateneo Nuovo, 1}
    \city{Milano}
    \country{Italy}
    \postcode{20126}
}
\author{Markus Zanker}
\email{mzanker@unibz.it}
\orcid{0000-0002-4805-5516}
\affiliation{%
    \institution{Free University of Bozen-Bolzano, Bolzano, Italy}
    \streetaddress{Piazza Università, 1}
    \city{}
    \country{}
    \postcode{39100}
}
\affiliation{%
    \institution{University of Klagenfurt, Klagenfurt, Austria}
    \streetaddress{Universitätsstraße 65-67}
    \city{}
    \country{}
    \postcode{9020}
}

\begin{abstract}
    Causality is receiving increasing attention in the Recommendation Systems (RSs) community, which has realised that RSs could greatly benefit from causality to transform accurate predictions into effective and explainable decisions. Indeed, the RS literature has repeatedly highlighted that, in real-world scenarios, recommendation algorithms suffer many types of biases since assumptions ensuring unbiasedness are likely not met. In this discussion paper, we formulate the RS problem in terms of causality, using potential outcomes and structural causal models, by giving formal definitions of the causal quantities to be estimated and a general causal graph to serve as a reference to foster future research and development. 
\end{abstract}


\maketitle

\input{sections/01_introduction}
\input{sections/02_rs_through_causal_lens}
\input{sections/03_conclusions}

\bibliographystyle{ACM-Reference-Format}
\bibliography{references}

\end{document}

%% file: sections/01_introduction.tex
\section{Introduction}
Predicting and deciding are two fundamentally different tasks. As described by the \textit{Ladder of Causation} \cite{pearl2018book}, a decision manipulates the system which can react to our decision, while a prediction does not affect the system in any manner: the system is eventually affected only when we exploit the prediction to make a decision. Overlooking this difference usually leads to biased predictions that, in turn, result in wrong decisions. The RSs community is facing several problems with biased estimates \cite{chen2023bias} to assess the effect of recommendations based on predictions. Indeed, according to \cite{adomavicius2005toward,ricci2011introduction}, the recommendation problem is usually framed as a prediction problem while, as pointed out in \cite{jeunen2022pessimistic,cavenaghi2024towards}, it is indeed a decision-making problem, since we have to decide which item(s) to recommend.

Furthermore, human beings are not interested in mere correlations but in understanding the actual causes of the effects, manipulating the world to achieve the desired outcome. In fact, scientists are familiar with the phrase: ``Correlation is not causation'', that is, for example, ``the rooster's crow is highly correlated with the sunrise; yet it does not cause the sunrise'' \cite{pearl2018book}. Indeed, under general conditions, machine learning approaches do not allow to state that $X$ is the cause of $Y$ but only that they are ``correlated'' to each other. 

\begin{figure*}
    \centering
    \includegraphics[width=0.7\textwidth]{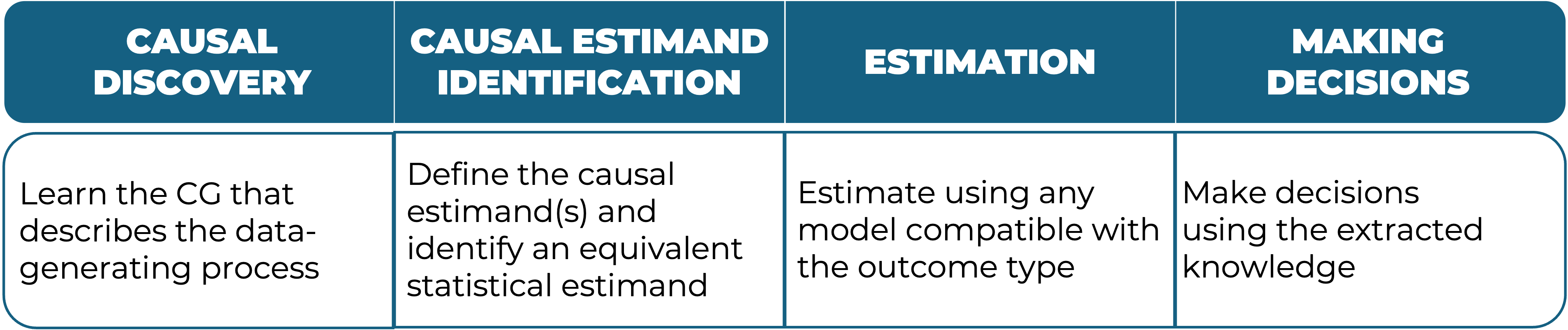}
    \caption{From available data and expert knowledge to making decisions.} 
    \label{fig:process}
\end{figure*}

This is why causality becomes important: we need a way to translate cause-and-effect relations and interventions on a system using a mathematical formulation. To this end, in \cite{cavenaghi2024towards}, we proposed a causal decision-making framework for RSs using Potential Outcomes (POs) \cite{rubin1974estimating} and Causal Graphs (CGs) \cite{pearl2009causality}. Using this framework, we introduce the process, illustrated in Figure \ref{fig:process}, which allows to make decisions by combining data and expert knowledge.

%% file: sections/02_rs_through_causal_lens.tex
\section{Causal Decision-Making}

\subsection{Causal Discovery}

The first step is to have a CG that describes the data-generating process of the system under study. The CG can be learned by combining observational data with experts' knowledge through a process called \textit{causal discovery}, which is enabled by several causal discovery algorithms \cite{vowels2021d,zanga2022survey}. While the CG must be learned in each scenario, we proposed a reference CG for RSs \cite{cavenaghi2024towards} to guide the construction of a CG for specific RSs problems as done in \cite{cavenaghi2023causal,cavenaghi2023analysis}.

\begin{figure}[ht]
    \centering
    \begin{tikzpicture}[node distance={13mm}, 
        main/.style = {circle, draw=black!60, line width=0.3mm, minimum size=6mm},
        policy_node/.style={diamond, draw=black!60, line width=0.3mm, minimum size=10mm},
        cluster_node/.style={rectangle, draw=black!60, line width=0.3mm, minimum size=6mm},
        ]
        \node[cluster_node] (1) {$\mathbf{U}$};
        \node[cluster_node] (2) [right= 15mm of 1] {$\mathbf{C}$};
        \node[policy_node] (3) [below of=1] {$\pi_x$};
        \node[main] (4) [below of=3] {$X$};
        \node[cluster_node] (5) [above right= 2.5mm and 5mm of 4] {$\mathbf{I}$};
        \node[main] (6) [right= 15mm of 4] {$Y$};

        \draw[-latex, line width=0.3mm] (1) -- (3);
        \draw[-latex, line width=0.3mm] (1) to [out=345, in=105, looseness=1.0] (6);
        \draw[-latex, line width=0.3mm] (2) -- (3);
        \draw[-latex, line width=0.3mm] (2) -- (6);
        \draw[-latex, line width=0.3mm] (3) -- (4);
        \draw[-latex, line width=0.3mm] (4) -- (5);
        \draw[-latex, line width=0.3mm] (4) -- (6);
        \draw[-latex, line width=0.3mm] (5) -- (6);
    \end{tikzpicture}
    \caption{Causal Graph for item ID-based recommendation.}
    \label{fig:general_causal_model_for_RSs}
\end{figure}
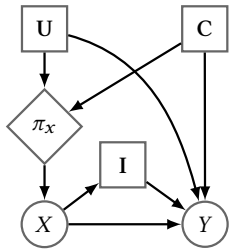

The problem of recommending a \textit{single} item with features $\mathbf{I}$ to a user $\mathbf{U}$ in context $\mathbf{C}$ is described by the CG of Figure \ref{fig:general_causal_model_for_RSs} where the node $X$ represents the action of recommending an item whose domain corresponds to the item set, i.e.,  $x \in dom(X)$. For example, in film recommendation, $dom(X)$ is the catalogue of the films and our recommendation $X$ is one of the films in the catalogue. To decide which item to recommend to the current user $\mathbf{U}$ in context $\mathbf{C}$, we use a policy $\pi_x$ based on user and context features. Once we decide which item $x$ to recommend, the corresponding item's features $\mathbf{I}$ are fixed and they mediate the effect of our recommendation $X$ on the user feedback $Y$ through the path $X \to \mathbf{I} \to Y$. For example, once we decide to recommend a film, its genre is fixed ($X \to genre$), and the genre (likely) affects a user's feedback ($genre \to Y$). It is worth noticing that not all the item's features have to influence the user's feedback, i.e., some item's features are not taken into account by the users. On the other hand, part of the effect of the recommendation $X$ on the user's feedback $Y$ may not be captured by the modelled features $\mathbf{I}$ and flows directly through the edge $X \to Y$. For example, if we are not able to model the film's popularity, its effect will flow through the edge $X \to Y$ as this feature is not modelled.

\subsection{Causal Estimand Identification}

To exploit the potential of causality, we should frame the quantity to estimate as a \textit{causal estimand} that encodes the notion of the causal effect of a variable (the cause) on another (the effect). We define it using the POs framework and the \textit{do-operator} \cite{bareinboim20211OP,glymour2016causal}, as $\mathbb{E}[Y|do(X=x),\mathbf{u},\mathbf{i},\mathbf{c}]$. This encodes the value of the expected feedback $Y$ given by the user $\mathbf{u}$ in context $\mathbf{c}$ when we recommend item $x$ with features $\mathbf{i}$. The difference that separates causal estimands from classical statistical estimands is the presence of the so-called \textit{do-operator}, denoted with $do(X = x)$, that defines the intervention of fixing the value of $X$ to $x$ for the whole population of users. In contrast, conditioning on $X=x$ means that $X$ takes a value $x$ naturally, which simply translates to focusing only on the sub-population where X has been observed to be equal to $x$. In a decision-making problem, such as RSs, we are interested in estimating causal estimands since we actively decide which item(s) to recommend.

However, expressions with the \textit{do-operator} can only be estimated in controlled experiments where the variables in the \textit{do-terms} can be appropriately controlled. To estimate a causal estimand using only observational data, it is necessary to remove the \textit{do-terms} and obtain an equivalent expression. To this end, the \textit{adjustment formula} estimator \cite{glymour2016causal} adopts a model-based approach to adjust for an \textit{adjustment set} $\mathbf{Z}$ and obtain a statistical estimand:

\begin{equation}\label{def:adjustment_formula}
    P(Y = y|do(X = x)) = \sum_\mathbf{z} P(Y = y|X = x, \mathbf{Z} = \mathbf{z})P(\mathbf{Z} = \mathbf{z})
\end{equation}

To \textit{identify} the variables that must be included in $\mathbf{Z}$, we can query the CG by evaluating an \textit{identification criterion}, such as the \textit{backdoor criterion} \cite{glymour2016causal}, \textit{frontdoor criterion} \cite{pearl2009causality} or \textit{do-calculus} \cite{bareinboim20211OP}. In particular, if no identification is possible with do-calculus, the causal effect is guaranteed to be \textit{unidentifiable}. Thus, every estimate of the causal estimand will be biased.

\subsection{Estimation}
Once we have proved the identifiability of the causal estimand, i.e., once we have shown that the causal estimand is equal to a statistical estimand, this can be estimated using classical statistical estimators. Any model that is compatible with the type of the outcome variable, e.g. linear regression for a continuous outcome or neural networks for non-linear relations, is suitable for this estimation. Clearly, the model should be chosen carefully for each problem by considering the data characteristics to avoid estimation errors.

\subsection{Making Decisions}

Finally, with the estimated causal effects, e.g., the effect of our recommendation on the user's propensity to click on the recommended item(s), we can decide which items to recommend. This could be done in different ways: (i) greedy, (ii) $\epsilon$-greedy and (iii) more sophisticated policies. In recent years, several works have exploited causality by linking it to \textit{Multi-Armed Bandit} (MAB) \cite{lattimore2016causal,lu2020regret,nair2021budgeted} and \textit{Reinforcement Learning} (RL) \cite{lu2021causal}. For example, in \cite{lee2018structural}, the authors define the notion of Possibly-Optimal Minimal Intervention Set with the idea of determining the minimum set of variables on which we should intervene to understand all the possible arms that are worth intervening on. Moreover, \cite{lee2019structural} extends the method by considering that some variables can not be manipulated. Using causality with RL, \cite{zhang2020designing,zhang2019near} approached the Dynamic Treatment Regimes problem with confounded observational dataset.

%% file: sections/03_conclusions.tex
\section{Conclusions}

In this paper, we proposed a causal view of the RS problem and highlighted the importance of framing the recommendation problem in terms of causality. The causality framework can, in our view, be considered as a single framework allowing researchers to wholistically define and address several problems widely acknowledged in the RSs community to bridge the gaps in future works.

However, we would like to stress that causality is not magic but ruthlessly honest and, differently from other approaches, it makes explicit assumptions, such as ignorability and unconfoundedness, leaving us with the burden of judging whether they are likely to be satisfied for the addressed context. Indeed, causality is not the sole ingredient to solve the RS problem while we are fully convinced that exploiting the body of knowledge generated over more than 30 years of research in RSs and users' behaviour remains fundamental.